**Title:** Enhancing the Properties of Plasma Activated Water using an Air Bubble Diffuser

*Authors*

Vikas Rathore[1,2], Nirav I. Jamnapara[1,2] and Sudhir Kumar Nema[1,2]

*Author affiliations*

[1]Institute for Plasma Research (IPR), Gandhinagar, Gujarat 382428, India

[2]Homi Bhabha National Institute, Training School Complex, Anushaktinagar, Mumbai 400094, India

*Author e-mails*

vikas.rathore@ipr.res.in

**Abstract**

In this study, the impact of an air bubbler on the properties of plasma-activated water (PAW) was investigated in different configurations. A pencil plasma jet (PPJ) using a dielectric barrier discharge (DBD) was used to prepare PAW. In one configuration, the air bubbler was fitted at the tip of the pencil plasma jet, causing the discharge gases to emerge as bubbles in the water. In another configuration, the water was agitated using the bubbler during plasma-water interaction. The plasma generated in the PPJ setup was a filamentary DBD micro-discharge. However, water agitation using the bubbler changed this filamentary DBD to a diffusive DBD, which showed higher discharge current and lower electrode voltage compared to filamentary

DBD. PAW produced using the fitted bubbler in the PPJ setup showed enhanced physicochemical properties, including $NO_2^-$, $NO_3^-$, dissolved $O_3$, and non-traceable $H_2O_2$ compared to PAW produced without a bubbler. Additionally, PAW produced using water agitation by the bubbler showed a substantial increase in physicochemical properties and reactive species concentrations. The PAW process parameters such as air flow rate, plasma-water treatment time, and plasma discharge power showed monotonically increasing properties of PAW. The maximum concentrations of $NO_2^-$, $NO_3^-$, $H_2O_2$, and dissolved $O_3$ (flow rate: 20 L min$^{-1}$, time: 5 min, power: 15 W) observed in this study were 0.334 g L$^{-1}$, 0.078 g L$^{-1}$, 0.045 g L$^{-1}$, and 0.016 g L$^{-1}$, respectively. This enhancement in the properties of PAW using a bubbler adds value to the current PAW technology in fields such as microbial inactivation, medicine, agriculture, and aquaculture, etc.

1. Introduction

Plasma Activated Water (PAW) technology has been proven to have potential in a wide range of applications over the last 15 years, from inactivating microbes to accelerating seed germination and plant growth. Some of the most recent applications include selective inactivation of cancer cells, virus inactivation, pest control, wound healing, and vaccine preparation. In the past decade, the most extensively studied applications of PAW have been in bacteria inactivation and food preservation. This is possible due to the various reactive species present in PAW (1-17).

PAW is defined as water that has been chemically modified as a result of exposure to plasma. Different gases produce different reactive species or radicals, with the major contributors to PAW chemistry being ·H, ·OH, O, N, $N_xO_y$ (NO, $NO_2$, $NO_3$, $N_2O_4$, $N_2O_5$, etc.), and $O_3$, etc. The dissolution of these reactive species in water results in a chemically modified water with a reduced pH, increased oxidizing tendency, and elevated electrical conductivity. The reduction in pH occurs due to the formation of nitrous and nitric acid in the water, while the increased oxidizing potential is a result of the combination of acidic radicals, peroxynitrite, hydrogen peroxide, and dissolved ozone, among others. The elevated electrical conductivity is due to the dissolution of inorganic ions such as $H^+$, $NO_2^-$, and $NO_3^-$, among others. The reactivity of PAW is determined based on the concentration of these reactive species. A high-reactive PAW has a high concentration of these reactive oxygen-nitrogen species (RONS) and low pH, high oxidizing tendency, and high electrical conductivity, while the opposite is true for a low-reactive PAW (4, 11, 12, 14, 16, 18-20).

Based on the reactivity of plasma-activated water (PAW), its applications have been categorized into two groups: high reactivity and low to moderate reactivity. High reactivity PAW is used for inactivating bacteria, fungi, viruses, and pests, while low to moderate reactivity PAW is used for food preservation, seed germination, and plant growth, as the high reactivity of PAW may damage these. The reactivity of PAW can be controlled by adjusting various process parameters, such as the type of plasma (gliding arc, spark, microwave, glow, filamentary dielectric barrier discharge, etc.), plasma-water treatment time, and plasma discharge power, etc. (2-4, 6, 9-11, 13, 15, 20-23)

Enhancing the dissolution of reactive species in water remains a challenge in the production of plasma-activated water (PAW). Despite high plasma-water interaction time and high plasma discharge power, poor dissolution of plasma reactive species and radicals in water results in low reactivity of the produced PAW. In recent years, researchers have attempted various configurations to optimize the reactive absorption of generated plasma species and radicals for water activation. One of the promising technologies in this field is plasma bubblers for PAW production, which increases the plasma residence time with water and reduces the loss of plasma species and radicals. The high surface-to-volume ratio of plasma bubbles is also believed to significantly enhance the reactivity of water. Previous literature has explored various techniques, including bubble-activated water, plasma-bubble column reactors, underwater bubbler discharge, microplasma bubbles, nanosecond pulsed microbubble plasma reactors, and more. (13, 14, 18, 24, 25)

This present work aims to address the research gap and enhance the dissolution of plasma reactive species and radicals in water. Two different configurations of a bubble diffuser (bubbler) are used during PAW production. In the first configuration, the bubbler is fitted to a PPJ setup so that discharge gases are diffused in water in the form of bubbles. In the second configuration, the bubble diffuser is used to turbulate the water, increasing the humidity in the discharge air and trapping undissolved plasma species and radicals. The plasma generated in both configurations is characterized by studying the voltage-current waveform. The PAW characterization is performed by examining changes in various physicochemical properties and RONS concentrations.

2. Material and Methods

2.1 Pencil plasma jet setup and plasma characterization

A Pencil Plasma Jet (PPJ) was utilized for PAW generation. The PPJ setup followed a cylindrical co-axial dielectric barrier discharge geometry for air plasma generation. The PPJ was powered by a high voltage high frequency power supply with a variable voltage range of 0-10 kV and a variable frequency range of 10-40 kHz. Compressed air was fed into the PPJ setup from the top and the flow rate was controlled using a flow controller (air rotameter). The generated plasma was characterized using voltage-current waveform, and the plasma discharge power was calculated using a voltage-transport charge Lissajous figure. The generated air plasma species/radicals were identified by capturing emission photos using an optical fiber and an optical spectrophotometer. Further details regarding the PPJ setup, electrical characterization, and species/radical identification can be found in previously published works (2, 8, 23, 26). For PAW preparation, 40 ml of ultrapure milli-Q water (control) was placed in a 100 ml glass beaker.

2.1 Bubbler specification

The bubble diffuser (bubbler), model number SFH01 (inline oxygenation diffusion stone), was purchased from HENGKO Technology Co., Ltd. The bubble diffuser had a dimension of 0.5 inch × 0.6 inch (diameter × height) with a 2 µm pore size. It was made of stainless steel 316L with uniform apertures and tight particle bonding. The bubble diffuser produced micro bubbles in the size range of 0.1 – 120 µm in diameter.

2.2 Pencil plasma jet fitted with or without bubbler

The pencil plasma jet was fitted with a bubbler using a pneumatic pipe. For comparison, an equal length pneumatic pipe was placed in water under the same process parameters (as shown in Figure 1, Case I). The distance between the PPJ setup tip and the water surface was 5 cm. The air was fed from the top, and the flow rate was controlled using an air rotameter.

2.3 Disturbing the stationary phase of water (water turbulation) using a stirrer and bubbler

As the plasma is generated in the cylindrical co-axial assembly of the PPJ setup, it is pushed downwards towards the water surface. To expose the plasma to a new water surface, turbulation was created in the water using a magnetic stirrer (range: 0-1500 RPM) and a bubbler, as shown in Figure 1 (Case II). The air flow rate of the bubbler was fixed at 20 L min$^{-1}$.

2.4 Measuring the properties of PAW

The properties of PAW were analyzed by measuring its physicochemical properties and reactive oxygen-nitrogen species (RONS). The measured properties included pH, oxidation-reduction potential (ORP), total dissolved solids (TDS), and electrical conductivity (EC). The concentration of RONS in PAW was measured using semi-quantitative test strips or colorimeter test kits. UV-visible spectroscopy was used for a quantitative estimation. The RONS in PAW included $NO_3^-$ ions, $NO_2^-$ ions, $H_2O_2$, and dissolved $O_3$. The $NO_3^-$ ions and $NO_2^-$ ions concentration were measured using the UV screening method at 220 nm and the Griess reagent system at 540 nm, respectively. The $H_2O_2$ concentration was measured using the titanium oxysulfate method at 407 nm, and the dissolved $O_3$ was measured using the Indigo colorimetry method at 600 nm (2, 8, 23, 26).

2.5 Data analysis

All experiments in this investigation were performed at least three times (n≥3). The collected data was analyzed and presented in results plots as mean ± standard deviation (μ ± σ). Statistically significant differences in the results were determined using a one-way analysis of variance (ANOVA) with a p-value of 0.05, 0.005, or 0.001, followed by a post-hoc test using Fischer's Least Significant Difference (LSD).

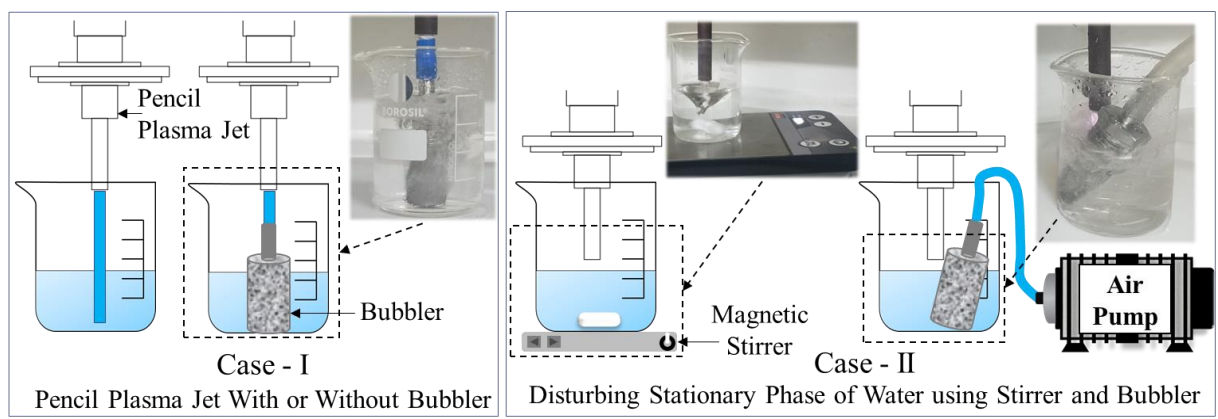

Figure 1. Schematic of Plasma Activated Water Production Using a Bubbler. Case I: Bubbler fitted with or without pencil plasma jet. Case II: Use of stirrer and bubbler in disturbing the stationary phase (turbulation or agitation) of water during plasma-water interaction.

3. Results and Discussion

3.1 Plasma diagnosis

The results of the plasma diagnosis in terms of electrical characterization and plasma reactive species identification are presented in Figure 2. Figure 2 (a, b) shows the voltage-current characteristics of air plasma during PAW production when the bubbler is fitted with a pencil plasma jet and when the bubbler creates turbulence in the stationary water phase. Two different types of dielectric barrier discharge are observed in both configurations. The PPJ fitted with the bubbler showed nanoscale discharge current filaments in each rising and falling cycle peak. These short-lived, nanosecond-range current filaments display multiple discharge current peaks over a continuous alternating current. The visual appearance of these electrical discharges looks like short-lived filaments to the naked eye. Additionally, the current and voltage are out of phase, as shown in Figure 2 (a). The combination of these current nanofilaments lies in the micro-range, and as such, this discharge is known as filamentary micro-discharge (27).

The voltage-current profile observed in PAW generation by turbulating water using a bubbler was substantially different from the bubbler fitted with PPJ. It showed a higher current and lower voltage drop across the PPJ setup. The peak-to-peak voltage in this configuration was 41.03% lower, and the current was 47.58% higher compared to the earlier configuration. Each rising and falling current half-cycle only showed a single major peak in this configuration, as shown in Figure 2 (b) (28). This characteristic implies the diffuse DBD discharge produced during PAW generation during water turbulence when using a bubbler. This discharge is relatively denser, uniform, louder, and brighter compared to filamentary DBD when viewed through the naked eye and examined in terms of noise level. The increasing current and reducing electrode voltage signifies the rapidly increasing electrical conductivity of the discharge gas. This is likely due to the water turbulence using the bubbler generating air-containing water bubbles that move upwards due to their low density and encounter the plasma region of the DBD filamentary discharge of the PPJ setup. This air-mixed bubble discharge in the plasma region increases the current density and reduces the electrode voltage, as shown in Figure 2 (b). The significance of these different discharges on the properties of PAW will be examined in upcoming sections.

The emission spectrum of air plasma is shown in Figure 2 (c). The main observed emission band peaks are $NO_\gamma$, $N_2$ second positive system (SPS), and $N_2^+$ first negative system (FNS), respectively, as shown in Figure 2 (c) (29). The intensity of N2 SPS is substantially higher compared to the other identified species in plasma, as shown in Figure 2 (c). These identified plasma species/radicals mainly participate in various dissociation and association reactions in the plasma afterglow region. The $N_2$, $O_2$, and $H_2O$ present in air are mainly ionized/dissociated by absorbing high energy from plasma or by impact ionization/dissociation by electrons or atoms/molecules. The dissociated products from plasma in the afterglow plasma-liquid interaction either undergo association reactions to form reaction products like $N_xO_y$, $H_2O_2$, $O_3$, etc., or remain in the dissociated form like ·OH, ·H, N, O, etc. (1, 4, 5, 7, 9, 15, 16, 19, 23, 26, 30, 31) These post-discharge plasma products interact with water to form various reactive species in water, as discussed in the next section.

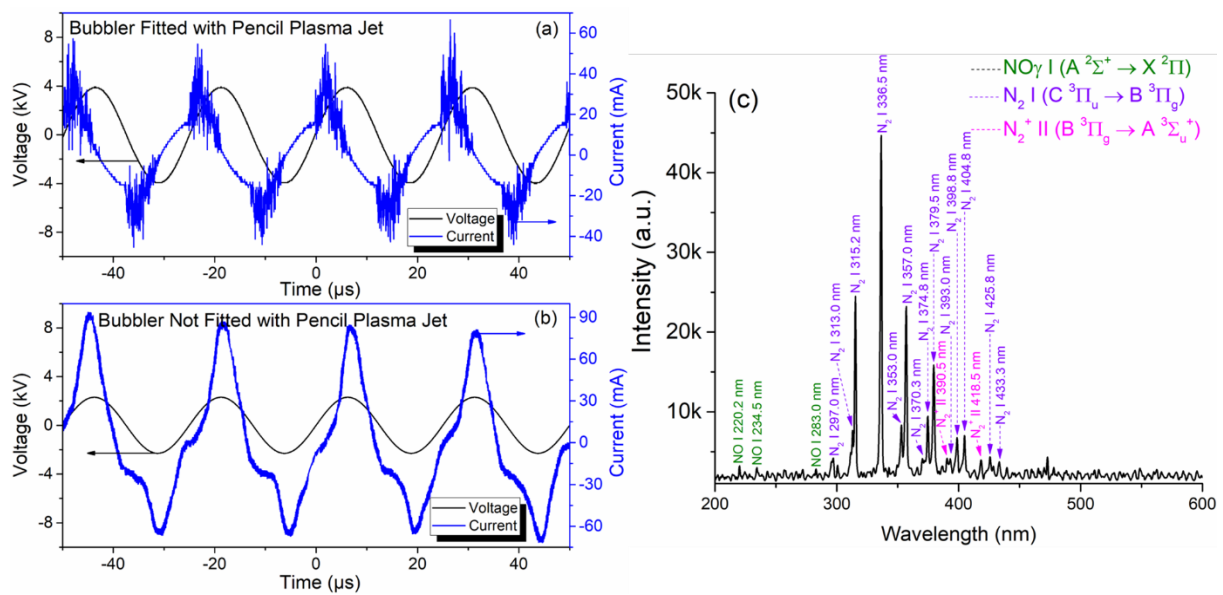

**Figure 2.** Characterization of plasma using voltage-current waveform and identification of plasma species using optical emission spectroscopy. (a) Bubbler fitted with pencil plasma jet and submerged in water, (b) Bubbler not fitted with pencil plasma jet and submerged in water (water turbulation), (c) Air plasma emission band peak lines.

### 3.2 Properties of PAW

In the previous section, the generation of plasma species, radicals, and high energy compounds/dissociated products at the plasma-liquid interface was discussed. These compounds and products interact and dissolve in water, leading to the formation of various reactive oxygen-nitrogen species (RONS), such as $NO_2^-$, $NO_3^-$, $N_2O_5$, $ONOO^-$, ·OH, $H_2O_2$, dissolved $O_3$, etc. The dissolution of these reactive species changes the physicochemical properties of water, such as decreasing the pH due to the formation of nitrous and nitric acids, and increasing the oxidizing potential of water due to the oxidizing nature of dissolved species. Moreover, since most of the dissolved species are inorganic, they increase the electrical conductivity of water after dissolution (1, 4, 5, 9, 10, 15, 17).

The current study investigates the impact of a metallic bubble diffuser on the properties of plasma activated water in two configurations. In the first configuration, the bubble was fitted directly to the pencil plasma jet, allowing discharge gases to directly enter the water as gas bubbles without loss (Case I). In the second configuration, the bubble was submerged in water without contact with the PPJ, and air was supplied externally to it for water turbulation or agitation (Case II). The effect of these configurations on the properties of PAW is studied in this work.

### 3.2.1 Role of bubbler fitted with PPJ on properties of PAW

The properties of PAW prepared using PPJ fitted with or without a bubbler, and with varying air flow rates, are shown in Figure 3. The PAW prepared using a bubbler fitted with PPJ showed better physicochemical properties compared to the non-bubbler PPJ setup. This enhancement was reflected in lower pH, higher electrical conductivity, increased concentrations of reactive nitrogen species ($NO_2^-$ ions and $NO_3^-$ ions), and dissolved $O_3$ (Figure 3 (a, d, e, f, h)). Bubbling the discharge gases in water enhances the $NO_2^-$ ions and $NO_3^-$ ions concentration by 109.8% and 34.1%, respectively, compared to the non-bubbler, at a higher gas flow rate of 20 L $h^{-1}$ (Figure 3 (e, f)). As a result, the pH of PAW (PPJ-bubbler) was 14.3% lower compared to PAW (PPJ-non-bubbler) (Figure 3 (a)). The TDS and EC of PAW (PPJ-bubbler) were 25.4% and 36.8%, respectively, higher than PAW (PPJ-non-bubbler) (Figure 3 (c, d)). The oxidizing potential (ORP) of PAW (PPJ-bubbler) and PAW (PPJ-non-bubbler) did not show any statistically significant difference ($p > 0.05$) with increasing flow rate. The PAW (PPJ-bubbler) did not show any observable concentration of $H_2O_2$, but it had a substantially high concentration of dissolved $O_3$ (22.5 mg/l) compared to PAW (PPJ-non-bubbler) (dissolved $O_3$ (max) = 2.7 mg $L^{-1}$) (Figure 3 (g, h)). The reason behind the non-traceable $H_2O_2$ in PAW (PPJ-bubbler) was the loss of OH radicals (half-life ~ $10^{-9}$ s) (32) while passing through the bubbler. As a result, the H2O2 concentration in PAW (PPJ-bubbler) was beyond the detection limit of the present work. The $H_2O_2$ and dissolved $O_3$ concentration in PAW (PPJ-non-bubbler) showed increasing and decreasing behavior with air flow rate. This was due to the fact that at high flow rates, the reactivity of PAW was high (low pH and high EC), which favored the reaction between $H_2O_2$, dissolved $O_3$, and $NO_2^-$ ions to produce stable $NO_3^-$ ions. Hence, the concentration of $H_2O_2$ and dissolved $O_3$ in PAW (PPJ-non-bubbler) decreased at high flow rates (4, 22, 31, 33).

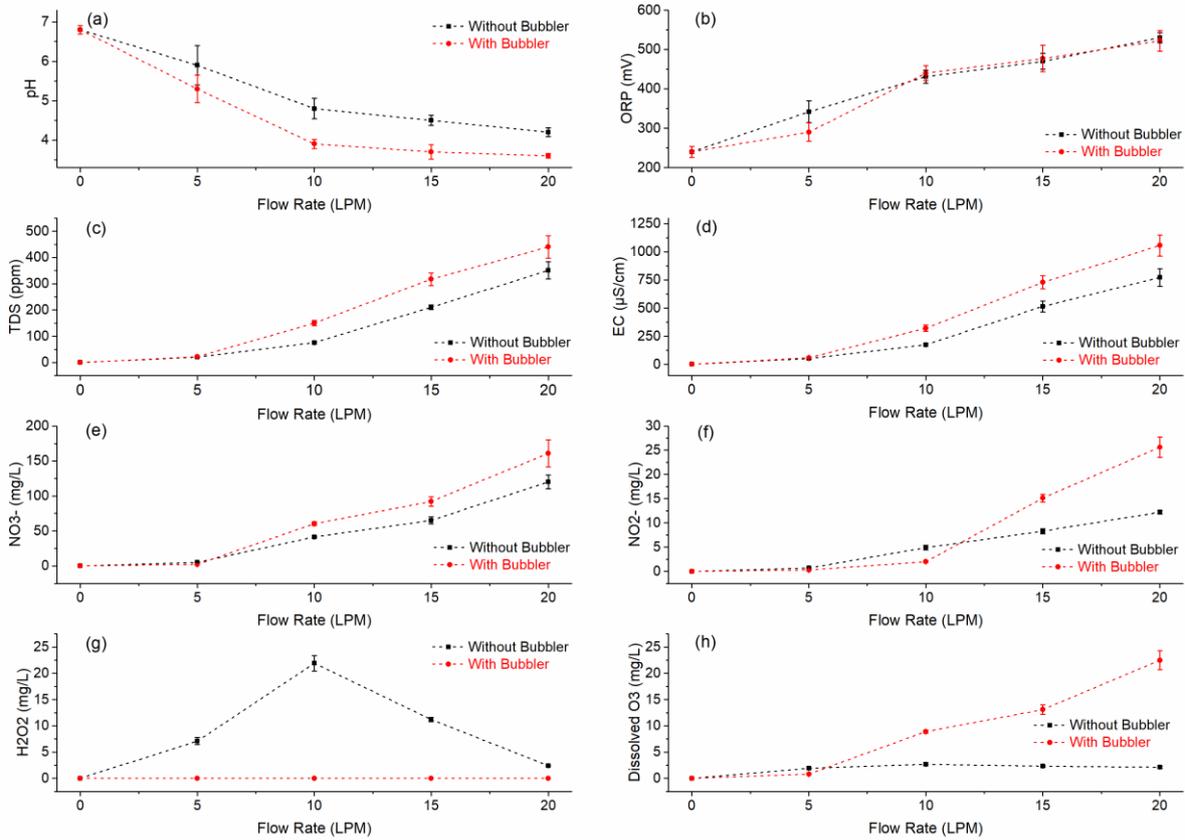

**Figure 3.** Properties of plasma activated water prepared using pencil plasma jet fitted with bubbler and without bubbler

3.2.2 Role of water turbulation (agitation) using bubbler and stirrer on properties of PAW

The effect of water turbulence using a bubbler and stirrer on the properties of PAW is presented in Figure 4. Figure 4 shows the changes in PAW properties when the flow rate of air through the PPJ and the stirring speed of water are varied. An increase in the flow rate improves the properties of PAW. For example, when the flow rate is increased from 5 to 20 L min$^{-1}$, the pH decreases from 4.2 to 3.1, while the ORP, TDS, EC, $NO_2^-$, $NO_3^-$, $H_2O_2$, and dissolved $O_3$ increase from 553 to 591 mV, 410 to 472 ppm, 861 to 1269.7 µS cm$^{-1}$, 161.2 to 185.4 mg L$^{-1}$, 30.6 to 65.8 mg L$^{-1}$, 2.7 to 35.7 mg L$^{-1}$, and 5.9 to 8.9 mg L$^{-1}$, respectively (Figure 4 (a1-a8)).

Moderate stirring (400 rpm and 600 rpm) in the range of 200 to 800 rpm also improves the properties of PAW as shown in Figure 4 (b1-b8). The optimum values of pH, ORP, TDS,

EC, and $NO_3^-$ ions occur at 600 rpm (Figure 4 (b1-b5)), while the $NO_2^-$, $H_2O_2$, and dissolved $O_3$ appear at 400 rpm (Figure 4 (b6-b8)).

The results of water turbulence using a bubbler are significantly higher compared to stirring at the same operating parameters. The observed difference in the highest concentration of $NO_3^-$, $NO_2^-$, $H_2O_2$ and dissolved $O_3$ in PAW prepared using a bubbler compared to PAW prepared using a stirrer was 75.4%, 702.4%, 1600%, and 368.3%, respectively.

This substantial increase in the concentration of reactive species in PAW prepared using water turbulence using a bubbler is due to changes in plasma discharge characteristics. The conversion of DBD filamentary micro-discharge to DBD diffusive discharge led to this improvement in the properties of PAW. The higher discharge current in water turbulence using a bubbler and brighter plasma (visual appearance) indicate the generation of a high concentration of plasma/species radicals that enhance the properties of PAW (28). In addition, water turbulence using a bubbler significantly enhances the humidity around the plasma column of the PPJ setup. The water containing air bubbles when in contact with the PPJ plasma column results in the transition of DBD filamentary discharge to DBD diffusive discharge.

This increase in humidity and discharge of bubbles in the plasma region enhances the ·OH radicals and $NO_x$ that dissolve in water during plasma-water interaction, thus increasing the $H_2O_2$ and RNS ($NO_2^-$ and $NO_3^-$ ions) concentration as shown in Figure 4.

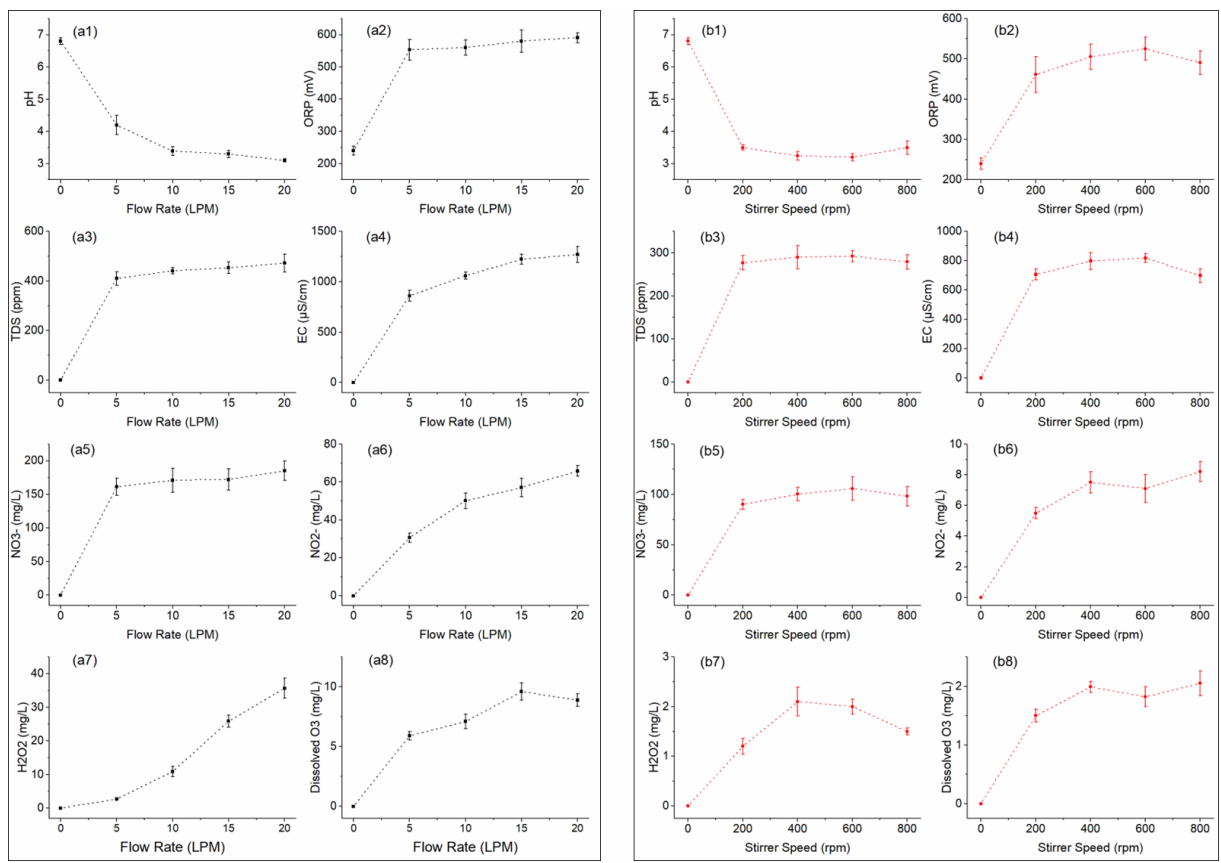

**Figure 4.** Properties of plasma activated water by water turbulation using bubbler and stirrer (varying air flow rate through PPJ setup and stirrer speed)

The use of water turbulation using a bubbler has shown a positive impact on enhancing the properties of PAW. Figure 4 only illustrates the role of air flow rate through the PPJ setup on the properties of PAW. Further investigation of other process parameters such as plasma-water treatment time and plasma discharge power on the properties of PAW during water turbulation using a bubbler is shown in Figure 5. Figure 5 shows that increasing plasma-water treatment time and plasma discharge power statistically significantly ($p < 0.001$) increases the properties of PAW. A longer plasma-water treatment time results in the increased formation of reactive species in PAW and improved physicochemical properties (Figure 5 (a1-a8)). A higher plasma discharge power signifies a higher concentration of plasma species/radicals (19, 20, 26) that when interacted and reacted with water generates a high concentration of RONS in PAW (Figure 5 (b1-b8)).

9 minutes of plasma-water treatment (water turbulation using a bubbler) reduces the pH of water from 6.8 to 2.9, enhances the ORP and EC from 240 mV to 610 mV and 1 µS cm$^{-1}$ to 1.67 mS cm$^{-1}$ (Figure 5 (a1, a3, a4)). At the same time, the observed concentration of $NO_3^-$, $NO_2^-$, and $H_2O_2$ and dissolved $O_3$ were 0.31 g L$^{-1}$, 0.08 g L$^{-1}$, 0.042 g L$^{-1}$, and 0.013 g L$^{-1}$, respectively. The 15 W plasma discharge power showed better properties of PAW compared to 9 minutes of plasma treatment time. However, these improved properties were less than or equal to ~10% (except for dissolved $O_3$). The reduction in pH of PAW – 15 W compared to PAW – 9 min was 6.9% and the increase in ORP and EC were 2.13% and 10.17% respectively. In addition, the improved concentration of $NO_3^-$, $H_2O_2$, and dissolved $O_3$ was 7.7%, 6.2%, and 21.7% respectively. This elevated concentration of dissolved $O_3$ showed a direct correlation between ozone production during DBD discharge with discharge power, which was in line with previously published literature. This enhancement in physicochemical properties and RONS concentration in PAW with increasing time and power is in line with previously published work of Jin et al. (20), Than et al. (10), Simon et al. (1), Raud et al. (31), Qiao et al. (7), Hoeben et al. (15), Pan et al.(4), Silsby et al. (16), etc.

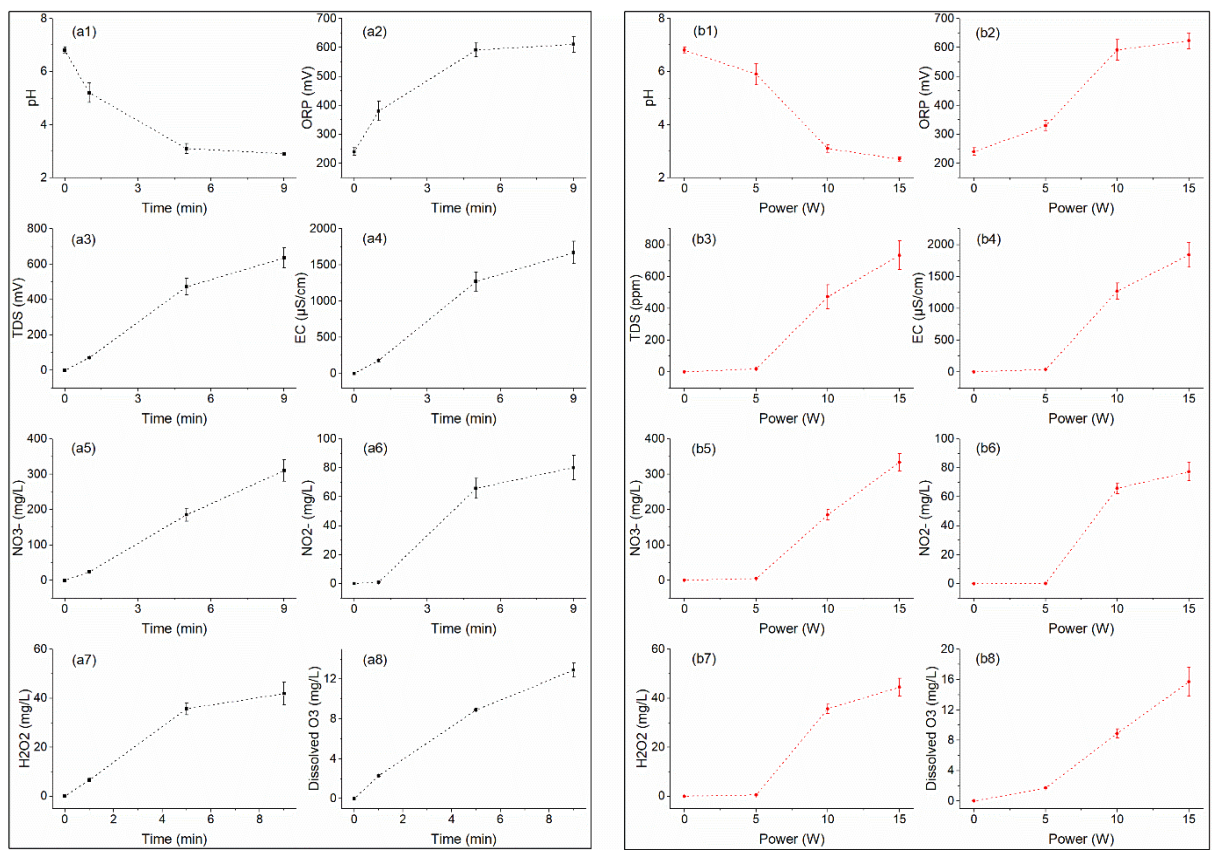

**Figure 5.** The variation in properties of plasma-activated water (PAW) generated through bubbler water turbulation, with varying plasma treatment time ($P_{constant}$ – 10 W) and plasma discharge power ($t_{constant}$ – 5 min).

The maximum energy efficiency of the various reactive oxygen and nitrogen species (RONS) present in PAW, within the given experimental constraints, were 10.691 g kW$^{-1}$h$^{-1}$ for NO$_3^-$ ions, 2.472 g kW$^{-1}$h$^{-1}$ for NO$_2^-$ ions, 1.424 g kW$^{-1}$h$^{-1}$ for H$_2$O$_2$, and 0.502 g kW$^{-1}$h$^{-1}$ for dissolved O$_3$. The total energy efficiency of RONS in PAW was 15.09 g kW$^{-1}$h$^{-1}$, which was 45.51% higher than the RONS energy efficiency reported by Man et al. (25). Man et al. (25) used a microbubble reactor for PAW production, but with a nanosecond pulsed microbubble and low water temperature (10°C) to enhance plasma species/radical dissolution. Our method did not use any additional cooling and still achieved higher RONS energy efficiency. Additionally, the observed RONS concentration was higher than that reported by Zhou et al. (13), Mai-Prochnow et al. (24), Feizollahi et al. (18) (without ice) who used underwater microplasma bubbles, atmospheric plasma bubbles, and bubble spark discharge for PAW generation. Zhou et al. (13) reported a RONS concentration of 1.5 mM (93 mg L$^{-1}$) $NO_3^-$ ions, 0.2 mM (9.2 mg L$^{-1}$) $NO_2^-$ ions, 1.3 mM (44.2 mg L$^{-1}$) $H_2O_2$, and 12 µM (0.56 mg L$^{-1}$) dissolved $O_3$ for a treatment time of 15 minutes, power of 40 W, and volume of 200 ml. These results signify the importance of diffusive DBD for PAW production that give value addition to current PAW technology to enhance the reactivity of PAW.    In summary, the bubbler fitted with a PPJ setup and water turbulation showed substantial improvement in PAW properties. However, the bubbler fitted with the PPJ setup is not recommended for $H_2O_2$ production as it resulted in non-detectable $H_2O_2$ concentration in PAW. The optimal configuration for PAW production, based on the results discussed, is water turbulation using a bubbler with high air flow rate, plasma-water treatment, and high plasma discharge power.

4. Conclusion

The present study describes a method for enhancing the properties of PAW (Plasma Activated Water) through the use of a bubble diffuser. Two different configurations of the bubbler were used in the production of PAW. In the first configuration, the bubbler was fitted with a PPJ setup, so the plasma discharge gases went through the bubbler and entered the water in the form of bubbles. This configuration was found to enhance the physicochemical properties and concentration of reactive species (except $H_2O_2$, which showed 0 mg L$^{-1}$) of PAW compared to the configuration without the bubbler, using similar parameters.

In the second configuration, the turbulence/agitation of the water was performed using the bubbler, causing the water mixed with air bubbles to come into contact with the plasma. This change of voltage-current profile from a filamentary discharge to a diffusive discharge led to substantial enhancements in the properties of PAW compared to other configurations. The study also showed that an increase in the process parameters of PAW production, such as air flow rate, plasma-water treatment time, and plasma discharge power, significantly increased the properties of PAW. In conclusion, the enhancement of the properties of PAW using the bubbler addresses current limitations of the PAW technology, such as short-term or limited antimicrobial efficacy, long exposure of PAW for food preservation and limited use of PAW as a pesticide. This method opens up new possibilities for PAW to be used in a wider range of applications due to enhanced reactivity.

*Acknowledgments*

This work was supported by the Department of Atomic Energy (Government of India) doctorate fellowship scheme (DDFS). The authors sincerely thank Mr. Chirayu Patil for providing constant support and useful suggestions during this work.

*Conflict of Interest*

The authors have no conflicts to disclose.

*Author Contributions*

V.R., N.I.J and S.K.N. contributed to conception and design of the study. Material preparation, data collection, and analysis were performed by V.R. The first draft of the manuscript was written by V.R., and all authors commented on previous versions of the manuscript. All authors read and approved the final manuscript.

*Data availability*

The data that support the findings of this study are available from the corresponding author upon reasonable request.